\documentclass[twocolumn]{revtex4}
\usepackage{graphicx}
\usepackage{subfigure}
\usepackage{ulem}

\makeatletter

\begin{document}

\title{Enhanced vaccine control of epidemics in adaptive networks}

\author{Leah B.~Shaw}

\affiliation{Department of Applied Science, College of William and Mary,
Williamsburg, VA 23187}

\author{Ira B.~Schwartz}

\affiliation{US Naval Research Laboratory, Code 6792, Nonlinear Systems
  Dynamics Section, Plasma Physics Division, Washington, DC 20375}

\begin{abstract}
We study vaccine control  for disease spread on
an adaptive network modeling disease
avoidance behavior. Control is implemented by  adding
Poisson distributed vaccination of susceptibles.  We show that vaccine control is much more effective in adaptive
networks than in static networks due to an interaction between the adaptive
network rewiring and the vaccine application. Disease extinction rates using
vaccination are computed, and orders of magnitude less vaccine application is needed to
drive the disease to extinction in an adaptive network than in a static
one.
\end{abstract}

\maketitle

Modeling the spread of epidemics on static networks is a well developed field,
but recent studies have begun to account for potentially time-varying network
topologies.  In particular, people may adjust their social behavior in
response to the threat of an epidemic.  Both an SIS
(susceptible-infected-susceptible) model \cite{GrossDB06} and SIRS
(susceptible-infected-recovered-susceptible) model \cite{ShawS08} have been
studied on an adaptive network in which non-infected nodes rewire their links
adaptively away from infected neighbors and toward other non-infected nodes.
Such adaptation typically increases the epidemic threshold and reduces the
number of infectious cases, and new disease dynamics and bifurcations are
observed.  Similar results are seen when the nodes reconnect to randomly
selected nodes elsewhere in the network \cite{Zanette2008,Risau-Gusman2009}. 
Because vaccines are available for many diseases, it is desirable to examine
the interplay between adaptively fluctuating social contacts and vaccination of susceptible individuals.

Almost all diseases exhibit randomness resulting in observed fluctuations, as in \cite{BolkerGrenfell93,Bolker93,Patz02b,Rand-Wilson,BBS-PRL,AndersonBritton2000}.
 As diseases evolve in large populations,
there is the possibility of finite time extinction  \cite{AndersonBritton2000,Herwaarden1995,allen00,jacquez93,Elgart2004,Doering2005}.
Extinction occurs
where the number of infectives become so small that there is insufficient
transmission to keep the disease in its endemic state \cite{keeling04,verdaska05,bartlett49}. Fluctuations cause the disease free equilibrium (DFE) to be reached in a finite time.  Such an extinction process occurs even when
the DFE is unstable.  Populations based upon adaptive networks
further complicate the problem, since social dynamical situations,
such as disease avoidance strategies, can cause the endemic and DFE to
be bistable \cite{ShawS08}. 

A major characteristic of fluctuation-induced extinction in stochastic
models for globally connected large populations is the extinction
rate. Viewing disease fade-out as coming from systems far from thermal
    equilibrium, finite population extinction rate laws have been derived in SIS 
  \cite{Doering2005,schwartzbdl2009} and SIR \cite{Kamenev2008} models. 
In  contrast to vaccine strategies  which stabilize the DFE  \cite{AMbook,d'Onofrio2002,Gao2007,Shulgin1998,Stone2000,wang2009}, periodic pulsed vaccination
was generalized to a random Poisson  strategy, which exponentially enhances
the rate of extinction \cite{dykmanschwartz08}. 

Other vaccine strategies have been examined in a variety of static  network geometries.  Targeting of high degree
nodes is widely recognized as more effective than random
vaccination, including in  scale free networks
  \cite{Pastor-Satorras2002,Dezso2002}, small world networks
  \cite{Zanette2002}, and more realistic social
network geometries \cite{Miller2007}.
Because targeting the highest degree
nodes requires full knowledge of the network geometry, other strategies based on local knowledge have
been developed.  For example, vaccinating a random acquaintance of a randomly
selected node tends to favor high degree nodes and is again more effective
than random vaccination \cite{Cohen2003}.  When vaccine is very limited,
outbreaks can be minimized by fragmenting the network via a graph partitioning
strategy which requires less vaccine than targeting high degree nodes
\cite{Chen2008}.  Here we will use a random vaccination strategy and find that
in conjunction with adaptive rewiring, it is extremely effective. 

We create a new  model by modifying an SIS model \cite{GrossDB06}, adding a vaccine class (V).  Individuals are connected in a network, with non-infected nodes rewiring adaptively to reduce connections with infected nodes.

The transition probabilities are as follows. A susceptible node becomes infected with rate $p N_{I,\text{nbr}}$, where
$N_{I,\text{nbr}}$ is its number of infected neighbors. An infected node
recovers (to susceptible) with rate $r$.  Vaccination occurs in Poisson-distributed pulses with frequency $\nu$, and each pulse vaccinates a fraction $A$ of the susceptible nodes \cite{NationalImmDays}.   We assume that the vaccine is not permanent, so a vaccinated node becomes susceptible again with rate $q$, the resusceptibility rate.

While the epidemic spreads, the network is rewired adaptively.  Because the
vaccine wears off, we assume that vaccinated nodes are uncertain of their
infection status and thus rewire in the same way as susceptibles.  If a link
connects a non-infected node to an infected node, that link is rewired with
rate $w$ to connect the non-infected node to another randomly selected
non-infected node. Self links and multiple links between nodes are excluded. 

We  simulate the full adaptive network via Monte Carlo simulation in
  a similar fashion to \cite{ShawS08}. Vaccination events occur with average rate $\nu$, and in each event $A N_S$ susceptibles are selected
randomly for vaccination, where $N_S$ is the total number of susceptibles.
Results are presented here for  $N=10^4$
nodes and $K=10^5$ links.

 We have developed a mean field theory
for the dynamics of nodes and links following \cite{GrossDB06,ShawS08} .
$P_A$ denotes the probability of a node to be in state $A$, where $A$ is
either S, I, or V.  $P_{AB}$ denotes the probability that a randomly selected
link connects a node in state $A$ to a node in state $B$. $\eta=\eta(t)$ is
the Poisson distributed  vaccination with  mean frequency $\nu$ and amplitude $A$.   We obtain the following set of stochastic differential equations for the nodes:
\begin{eqnarray*}
\label{eq:mf_node}
\dot{P}_S &=& r P_I -p \textstyle{\frac{K}{N}} P_{SI} -\eta(t) P_S+q P_V \\
\dot{P}_I &=& p \textstyle{\frac{K}{N}} P_{SI}-r P_I \\
\dot{P}_V &=& \eta(t) P_S -q P_V
\end{eqnarray*}
and for the links:
\begin{eqnarray*}
\dot{P}_{SS} &=& r P_{SI}  -2p \textstyle{\frac{K}{N}} \frac{P_{SS}P_{SI}}{P_S}-2 \eta(t) P_{SS} +q P_{SV} \nonumber \\
& & -\eta^2(t) P_{SS} +w \textstyle{\frac{P_S}{P_S+P_V}} P_{SI}\\
\dot{P}_{SI} &=& 2p\textstyle{\frac{K}{N}} \frac{P_{SS}P_{SI}}{P_S} -p \left( P_{SI} + \textstyle{\frac{K}{N}} \frac{ P_{SI}^2}{P_S} \right)  -r P_{SI} \nonumber \\
& &+2rP_{II}+qP_{IV}-\eta(t) P_{SI} -w P_{SI} \\
\dot{P}_{SV} &=& r P_{IV} -p \textstyle{\frac{K}{N}} \frac{P_{SI} P_{SV}}{P_S} +2q P_{VV}-qP_{SV}   -\eta(t) P_{SV} \\
& & + 2 \eta(t) P_{SS}+w \textstyle{\frac{P_V}{P_S+P_V}} P_{SI}
+w \textstyle{\frac{P_S}{P_S+P_V}} P_{IV} \\
\dot{P}_{II} &=& p \left( P_{SI} + \textstyle{\frac{K}{N}} \frac{ P_{SI}^2}{P_S} \right) - 2r P_{II} \\
\dot{P}_{IV} &=& p \textstyle{\frac{K}{N}} \frac{P_{SI} P_{SV}}{P_S}  -r P_{IV}-q P_{IV} +\eta(t) P_{SI}  -w P_{IV} \\
\dot{P}_{VV} &=& \eta(t) P_{SV} -2q P_{VV} +\eta^2(t) P_{SS} +w \textstyle{\frac{P_V}{P_S+P_V}} P_{IV}
\label{eq:mf_links}
\end{eqnarray*}
The mean field equations are analyzed as a stochastic system with random $\eta(t)$ as above, as well
as for a deterministic system using the mean vaccine rate $\left<\eta(t)\right>=\nu A$.

\begin{figure}[tbp]
\includegraphics[width=3in,keepaspectratio]{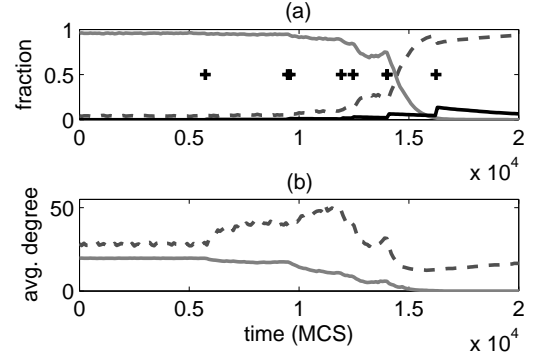}
\caption{Sample time series.  a) Node fractions.  Solid gray curve: infectives; dashed dark gray curve: susceptibles; solid black curve: vaccinated.  `+' symbols indicates times of vaccine pulses.  b) Average degree by node class.  Solid gray curve: infectives; dashed dark gray curve: susceptibles. $p=0.003$, $r=0.002$, $q=0.0002$, $A=0.1$, $\nu=0.0005$, $w=0.04$.  }
\label{fig:timeseries}
\end{figure}

Sample time series are shown in Figure \ref{fig:timeseries} for a run in which the epidemic became extinct.  Because we use proportional vaccination, as vaccination begins to lower the number of infectives and the number of susceptibles increases, subsequent vaccine pulses vaccinate a larger number of nodes.

We studied the longtime behavior for the case of longer lived endemic
states.  The deterministic mean field model predicts stable steady state
dynamics for the static network, even when vaccine is applied.  Small regions
of oscillatory behavior have been predicted for the adaptive network without
vaccination \cite{GrossDB06}, and these regions are predicted to be much
larger when vaccination is applied.  For the resusceptibility $q$ used here,
oscillations are generic \cite{note1}.

\begin{figure*}[tbp]
\centering
\subfiguretopcaptrue
\subfigure[]{
\includegraphics[width=2.7in,keepaspectratio]{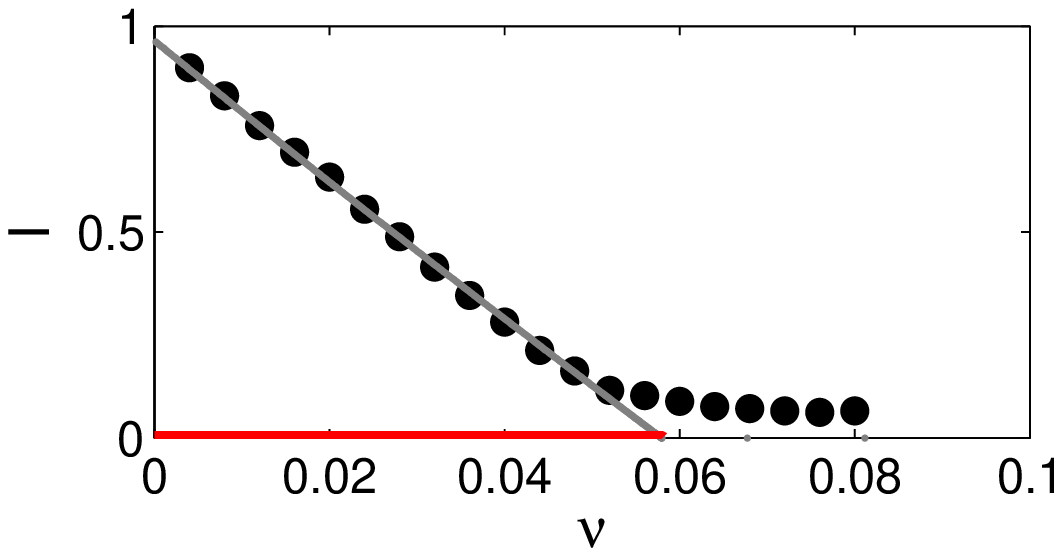}
\label{fig:w0MF}
}
\subfigure[]{
\includegraphics[width=2.7in,keepaspectratio]{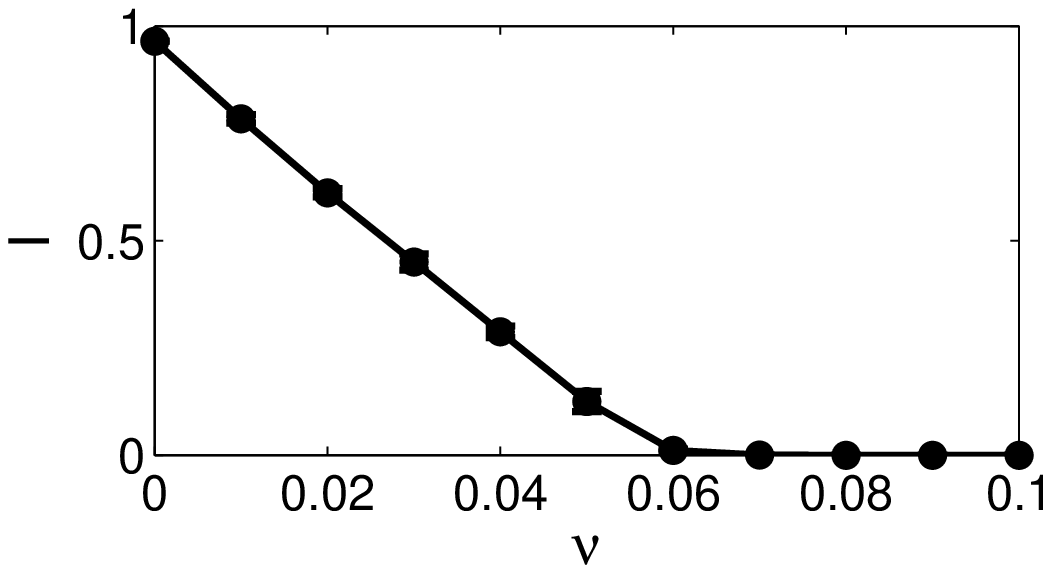}
\label{fig:w0MC}
}
\subfigure[]{
\includegraphics[width=2.7in,keepaspectratio]{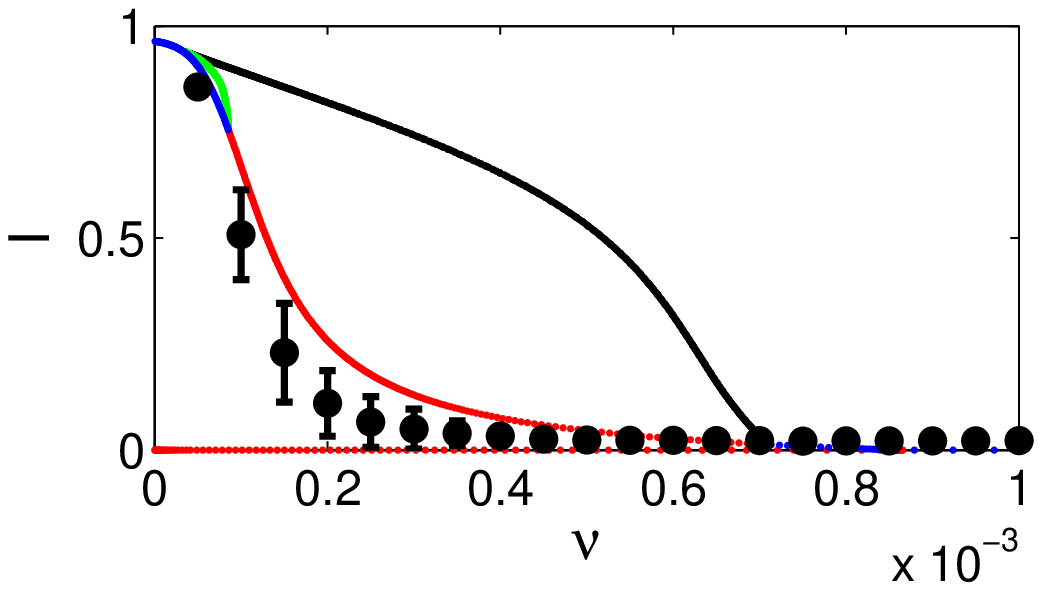}
\label{fig:w04MF}
}
\subfigure[]{
\includegraphics[width=2.7in,keepaspectratio]{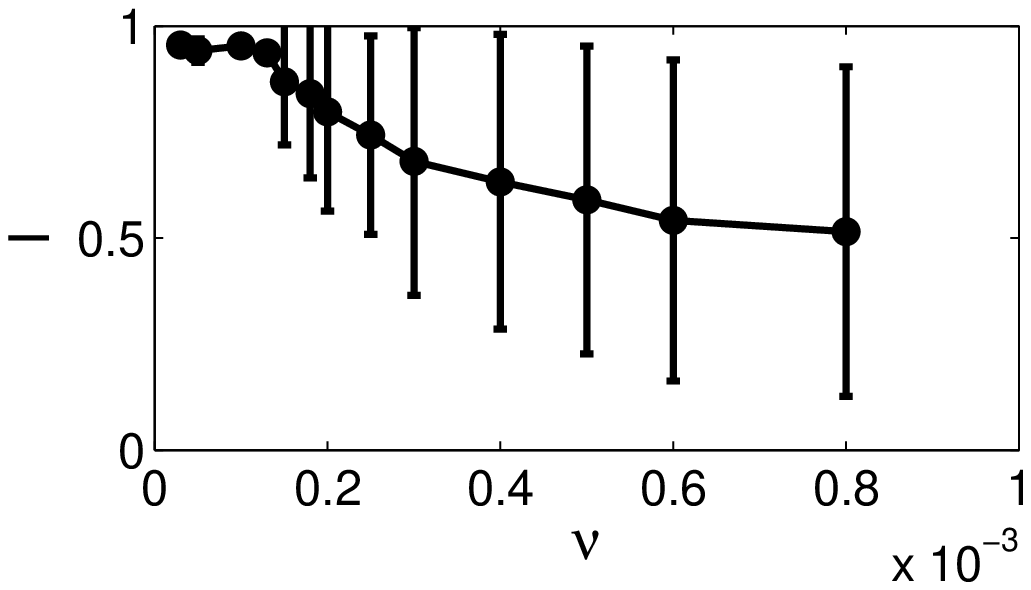}
\label{fig:w04MC}
}
\label{fig:subfigureExample}
\caption[]{Average infectives versus vaccine frequency $\nu$ for $q=0.0002, r-0.002, p=0.003, A=0.1$.  \subref{fig:w0MF} Mean field, $w=0$.  Gray and red curves, respectively, are stable and unstable branches of the deterministic mean field.  Symbols are stochastic mean field.
\subref{fig:w0MC} Full system, $w=0$.  Curve is to guide the eyes.
\subref{fig:w04MF} Mean field, $w=0.04$.  Blue curve: stable steady state, red curve: unstable steady state, black curve: stable periodic orbit, green curve: unstable periodic orbit for deterministic model.  Symbols are stochastic mean field.
\subref{fig:w04MC} Full system, $w=0.04$. (Color online.)
}
\end{figure*}

Figures \ref{fig:w0MF}-\ref{fig:w04MC} compare the dependence of the mean
infective level on the vaccine frequency for static and adaptive networks \cite{note2}.
For the full static (Erd\H{o}s-R\'{e}nyi) network system (Fig.~\ref{fig:w0MC}),
at each vaccine frequency the steady state mean infectives were computed over  ten network geometries. For the stochastic mean field (Fig.\ref{fig:w0MF}), 100 runs were done to extinction ($P_I<10^{-12}$)  and the
means of the time series were computed.  The mean field and full model are in excellent agreement.

For the full system with rewiring (Fig.~\ref{fig:w04MC}), infectives were
averaged over $8 \times 10^4$ MCS for a single adaptive network for the
smaller vaccine frequencies ($\nu \leq 0.00015$), and error bars are the
standard deviation. For larger $\nu$ values the oscillations led to rapid die out, so the
infectives were instead averaged over 100 runs computed to extinction, where the vaccine was turned on at
time 0.  Error bars are the
standard deviation of all the time points.  This averaging includes transients
 but serves to illustrate the
decreasing infective levels and large fluctuations due to oscillations as $\nu$ is
increased.    The stochastic mean field (Fig.~\ref{fig:w04MF}) was computed as in Fig.~\ref{fig:w0MF}, and error bars are the standard deviations of the
means.
The vaccination frequency required to significantly lower the infected fraction in the adaptive network is much smaller than for the static network due to the interaction of vaccination and rewiring.  The mean field model accurately predicts the order of magnitude of vaccine required and the presence of oscillations.

\begin{figure}[tbp]
\includegraphics[width=3in,keepaspectratio]{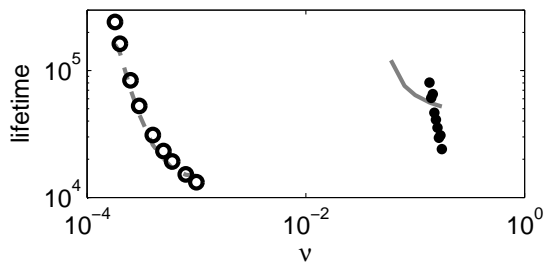}
\caption{Dependence of endemic state average lifetime on vaccine frequency $\nu$. Black open circles and dashed gray curve are full system and mean field, respectively, with rewiring ($w=0.04$). Black closed circles and solid gray curve are full system and mean field, respectively, with no rewiring.  $p=0.003$, $r=0.002$, $q=0.0002$, $A=0.1$.  }
\label{fig:lifetimes}
\end{figure}

Figure \ref{fig:lifetimes} shows the dependence of the endemic state lifetime
on the vaccine frequency for static (Erd\H{o}s-R\'{e}nyi) and adaptive networks.  About two orders
of magnitude less vaccine is needed in the adaptive case to significantly
reduce the lifetime of the endemic state.  For the full system, each point
was obtained by averaging 100 runs for which the initial condition was the
vaccine-free steady state and the vaccine was turned on at time zero.  Mean
field results were obtained similarly, but the time to extinction was
  computed using a threshold for $P_I$ of $10^{-12}$.  

\begin{figure}[tbp]
\includegraphics[width=2in,keepaspectratio]{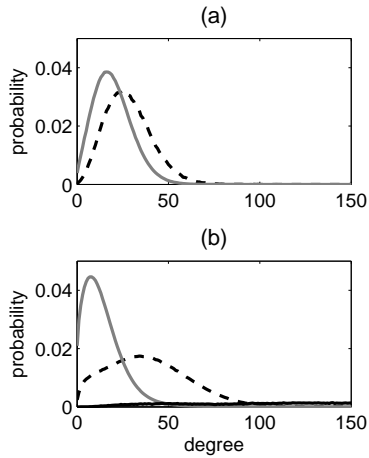}
\caption{Degree distributions from Monte Carlo simulation for $p=0.002$, $r=0.002$, $w=0.04$.  (a) No vaccination.  (b) With vaccination:  $\nu=0.00015$, $A=0.1$, $q=0.0002$.  Degree distribution of vaccinated nodes is very broad, extending beyond the figure domain, with an average degree of 579.  Solid gray:  infectives; dashed:  susceptibles; solid black:  vaccinated.}
\label{fig:degdistr}
\end{figure}

To explain the efficacy of vaccination in the adaptive network, we 
examine the network structure in more detail, particularly the degree.  Steady state degree
distributions for an adaptive network with and without vaccination are shown
in Figure \ref{fig:degdistr}.  Results without vaccination were obtained similarly to those in \cite{ShawS08}.
Results with vaccination were computed likewise but averaging over 9 runs to obtain better statistics for the vaccinated nodes
which are present at very low levels (0.6\% of the nodes).  As shown in
Fig.~\ref{fig:degdistr}(a), susceptibles in the adaptive network have a higher
average degree than infectives due to rewiring.  This is also apparent at the beginning of the
degree time series for susceptibles and infectives in
Fig.~\ref{fig:timeseries}(b).  Because the susceptibles typically have higher
degree, random vaccination of susceptible nodes favors the higher degree nodes
in the network and is therefore expected to be effective, as in previous
studies of targeted vaccination
\cite{Pastor-Satorras2002,Zanette2002,Dezso2002}.  For a static network, in
contrast, the high degree nodes are most likely to be infected and thus will
rarely be selected for vaccination. 

  The vaccination level used in Fig.~\ref{fig:degdistr}(b)  is small enough that it only slightly lowers the number of infected nodes.  However, when vaccination occurs the rewiring becomes more effective because of the presence of ``safe'' nodes to wire to.  The average infective degree is substantially decreased, and the susceptible degree is substantially increased.  The vaccinated nodes have a broad degree distribution with a very high average degree.  For the parameter values in Figure \ref{fig:degdistr}(b), vaccinated nodes participate in links as frequently as susceptibles, even though there are two orders of magnitude fewer vaccinated than susceptibles.

In this model, we randomly vaccinated a fixed fraction $A$ of the susceptible
nodes during each vaccination event.  Information about the network structure
was not required.  This corresponds to the case where vaccine is inexpensive
and anyone who is not infected can be vaccinated.  If the vaccine were in
limited supply, it would be desirable to use some knowledge about the social
network structure (such as the degrees) to target the vaccine more carefully.
In that case, the vaccine pulses might be a fixed number of susceptibles
rather than a fixed fraction. However, further study is needed to determine the optimal vaccination strategy.  Another extension would be to relax the assumption that nodes have full knowledge of others' disease status
\cite{Zanette2008,Risau-Gusman2009}.

In summary, we studied the effect of Poisson  vaccination
on epidemic spread in an adaptive network.  Vaccination was  far
more effective (by two orders of magnitude for the parameters chosen) in an adaptive network than a static one due to the interaction
of vaccination and rewiring.  The network adaptation led to a higher average
degree for susceptible nodes, and random vaccination of susceptibles exploited
this heterogeneity.  The adaptive rewiring then became more effective because
of the presence of ``safe'' vaccinated nodes to rewire to. 

LBS was supported by the Jeffress Memorial Trust, Army Research Office, and
Air Force Office of Scientific Research.  IBS was supported by the Office of
Naval Research and the Air Force of Scientific Research.


\begin{thebibliography}{36}
\expandafter\ifx\csname natexlab\endcsname\relax\def\natexlab#1{#1}\fi
\expandafter\ifx\csname bibnamefont\endcsname\relax
  \def\bibnamefont#1{#1}\fi
\expandafter\ifx\csname bibfnamefont\endcsname\relax
  \def\bibfnamefont#1{#1}\fi
\expandafter\ifx\csname citenamefont\endcsname\relax
  \def\citenamefont#1{#1}\fi
\expandafter\ifx\csname url\endcsname\relax
  \def\url#1{\texttt{#1}}\fi
\expandafter\ifx\csname urlprefix\endcsname\relax\def\urlprefix{URL }\fi
\providecommand{\bibinfo}[2]{#2}
\providecommand{\eprint}[2][]{\url{#2}}

\bibitem[{\citenamefont{Gross et~al.}(2006)\citenamefont{Gross, D'Lima, and
  Blasius}}]{GrossDB06}
\bibinfo{author}{\bibfnamefont{T.}~\bibnamefont{Gross}},
  \bibinfo{author}{\bibfnamefont{C.~J.~D.} \bibnamefont{D'Lima}},
  \bibnamefont{and} \bibinfo{author}{\bibfnamefont{B.}~\bibnamefont{Blasius}},
  \bibinfo{journal}{Phys.~Rev.~Lett.} \textbf{\bibinfo{volume}{96}},
  \bibinfo{pages}{208701} (\bibinfo{year}{2006}).

\bibitem[{\citenamefont{Shaw and Schwartz}(2008)}]{ShawS08}
\bibinfo{author}{\bibfnamefont{L.~B.} \bibnamefont{Shaw}} \bibnamefont{and}
  \bibinfo{author}{\bibfnamefont{I.}~\bibnamefont{Schwartz}},
  \bibinfo{journal}{Phys.~Rev.~E} \textbf{\bibinfo{volume}{77}},
  \bibinfo{pages}{066101} (\bibinfo{year}{2008}).

\bibitem[{\citenamefont{Zanette and Risau-Gusm\'{a}n}(2008)}]{Zanette2008}
\bibinfo{author}{\bibfnamefont{D.~H.} \bibnamefont{Zanette}} \bibnamefont{and}
  \bibinfo{author}{\bibfnamefont{S.}~\bibnamefont{Risau-Gusm\'{a}n}},
  \bibinfo{journal}{J. Biol. Phys.} \textbf{\bibinfo{volume}{34}},
  \bibinfo{pages}{135} (\bibinfo{year}{2008}).

\bibitem[{\citenamefont{Risau-Gusm\'{a}n and Zanette}(2009)}]{Risau-Gusman2009}
\bibinfo{author}{\bibfnamefont{S.}~\bibnamefont{Risau-Gusm\'{a}n}}
  \bibnamefont{and} \bibinfo{author}{\bibfnamefont{D.~H.}
  \bibnamefont{Zanette}}, \bibinfo{journal}{J.~Theor.~Biol.}
  \textbf{\bibinfo{volume}{257}}, \bibinfo{pages}{52} (\bibinfo{year}{2009}).

\bibitem[{\citenamefont{Bolker and Grenfell}(1993)}]{BolkerGrenfell93}
\bibinfo{author}{\bibfnamefont{B.~M.} \bibnamefont{Bolker}} \bibnamefont{and}
  \bibinfo{author}{\bibfnamefont{B.~T.} \bibnamefont{Grenfell}},
  \bibinfo{journal}{Proc. Roy. Soc. Lond. B} \textbf{\bibinfo{volume}{251}},
  \bibinfo{pages}{75} (\bibinfo{year}{1993}).

\bibitem[{\citenamefont{Bolker}(1993)}]{Bolker93}
\bibinfo{author}{\bibfnamefont{B.~M.} \bibnamefont{Bolker}},
  \bibinfo{journal}{IMA J. Math. Appl. Med.} \textbf{\bibinfo{volume}{10}},
  \bibinfo{pages}{83} (\bibinfo{year}{1993}).

\bibitem[{\citenamefont{Patz}(2002)}]{Patz02b}
\bibinfo{author}{\bibfnamefont{J.}~\bibnamefont{Patz}}, \bibinfo{journal}{Proc.
  Natl. Acad. Sci.} \textbf{\bibinfo{volume}{99}}, \bibinfo{pages}{12506}
  (\bibinfo{year}{2002}).

\bibitem[{\citenamefont{Rand and Wilson}(1991)}]{Rand-Wilson}
\bibinfo{author}{\bibfnamefont{D.}~\bibnamefont{Rand}} \bibnamefont{and}
  \bibinfo{author}{\bibfnamefont{H.}~\bibnamefont{Wilson}},
  \bibinfo{journal}{Proc. Roy. Soc. Lond. B} \textbf{\bibinfo{volume}{246}},
  \bibinfo{pages}{179} (\bibinfo{year}{1991}).

\bibitem[{\citenamefont{Billings et~al.}(2002)\citenamefont{Billings, Bollt,
  and Schwartz}}]{BBS-PRL}
\bibinfo{author}{\bibfnamefont{L.}~\bibnamefont{Billings}},
  \bibinfo{author}{\bibfnamefont{E.}~\bibnamefont{Bollt}}, \bibnamefont{and}
  \bibinfo{author}{\bibfnamefont{I.}~\bibnamefont{Schwartz}},
  \bibinfo{journal}{Phys.~Rev.~Lett.} \textbf{\bibinfo{volume}{88}},
  \bibinfo{pages}{234101} (\bibinfo{year}{2002}).

\bibitem[{\citenamefont{Andersson and Britton}(2000)}]{AndersonBritton2000}
\bibinfo{author}{\bibfnamefont{H.}~\bibnamefont{Andersson}} \bibnamefont{and}
  \bibinfo{author}{\bibfnamefont{T.}~\bibnamefont{Britton}},
  \bibinfo{journal}{J. Math. Biol.} \textbf{\bibinfo{volume}{41}},
  \bibinfo{pages}{559} (\bibinfo{year}{2000}).

\bibitem[{\citenamefont{van Herwaarden and Grasman}(1995)}]{Herwaarden1995}
\bibinfo{author}{\bibfnamefont{O.~A.} \bibnamefont{van Herwaarden}}
  \bibnamefont{and} \bibinfo{author}{\bibfnamefont{J.}~\bibnamefont{Grasman}},
  \bibinfo{journal}{J. Math. Biology} \textbf{\bibinfo{volume}{33}},
  \bibinfo{pages}{581} (\bibinfo{year}{1995}).

\bibitem[{\citenamefont{Allen and Burgin}(2000)}]{allen00}
\bibinfo{author}{\bibfnamefont{L.}~\bibnamefont{Allen}} \bibnamefont{and}
  \bibinfo{author}{\bibfnamefont{A.~M.} \bibnamefont{Burgin}},
  \bibinfo{journal}{Math. Biosci.} \textbf{\bibinfo{volume}{163}},
  \bibinfo{pages}{1} (\bibinfo{year}{2000}).

\bibitem[{\citenamefont{Jacquez and Simon}(1993)}]{jacquez93}
\bibinfo{author}{\bibfnamefont{J.~A.} \bibnamefont{Jacquez}} \bibnamefont{and}
  \bibinfo{author}{\bibfnamefont{C.~P.} \bibnamefont{Simon}},
  \bibinfo{journal}{Math. Biosci.} \textbf{\bibinfo{volume}{117}},
  \bibinfo{pages}{77} (\bibinfo{year}{1993}).

\bibitem[{\citenamefont{Elgart and Kamenev}(2004)}]{Elgart2004}
\bibinfo{author}{\bibfnamefont{V.}~\bibnamefont{Elgart}} \bibnamefont{and}
  \bibinfo{author}{\bibfnamefont{A.}~\bibnamefont{Kamenev}},
  \bibinfo{journal}{Phys.~Rev.~E} \textbf{\bibinfo{volume}{70}},
  \bibinfo{pages}{041106} (\bibinfo{year}{2004}).

\bibitem[{\citenamefont{Doering et~al.}(2005)\citenamefont{Doering, Sargsyan,
  and Sander}}]{Doering2005}
\bibinfo{author}{\bibfnamefont{C.~R.} \bibnamefont{Doering}},
  \bibinfo{author}{\bibfnamefont{K.~V.} \bibnamefont{Sargsyan}},
  \bibnamefont{and} \bibinfo{author}{\bibfnamefont{L.~M.}
  \bibnamefont{Sander}}, \bibinfo{journal}{Multiscale Modeling \& Simulation}
  \textbf{\bibinfo{volume}{3}}, \bibinfo{pages}{283} (\bibinfo{year}{2005}).

\bibitem[{\citenamefont{Keeling}(2004)}]{keeling04}
\bibinfo{author}{\bibfnamefont{M.~J.} \bibnamefont{Keeling}},
  \emph{\bibinfo{title}{Ecology, Genetics, and Evolution.}}
  (\bibinfo{publisher}{Elsevier}, \bibinfo{address}{New York},
  \bibinfo{year}{2004}).

\bibitem[{\citenamefont{Verdasca et~al.}(2005)}]{verdaska05}
\bibinfo{author}{\bibfnamefont{J.}~\bibnamefont{Verdasca}}
  \bibnamefont{et~al.}, \bibinfo{journal}{J. Theor. Bio.}
  \textbf{\bibinfo{volume}{233}}, \bibinfo{pages}{553} (\bibinfo{year}{2005}).

\bibitem[{\citenamefont{Bartlett}(1949)}]{bartlett49}
\bibinfo{author}{\bibfnamefont{M.~S.} \bibnamefont{Bartlett}},
  \bibinfo{journal}{J. Roy. Stat. Soc. B} \textbf{\bibinfo{volume}{11}},
  \bibinfo{pages}{211} (\bibinfo{year}{1949}).

\bibitem[{\citenamefont{Schwartz et~al.}({2009})\citenamefont{Schwartz,
  Billings, Dykman, and Landsman}}]{schwartzbdl2009}
\bibinfo{author}{\bibfnamefont{I.~B.} \bibnamefont{Schwartz}},
  \bibinfo{author}{\bibfnamefont{L.}~\bibnamefont{Billings}},
  \bibinfo{author}{\bibfnamefont{M.}~\bibnamefont{Dykman}}, \bibnamefont{and}
  \bibinfo{author}{\bibfnamefont{A.}~\bibnamefont{Landsman}},
  \bibinfo{journal}{{J. Stat. Mech.}}  \bibinfo{pages}{P01005}
  (\bibinfo{year}{{2009}}).

\bibitem[{\citenamefont{Kamenev and Meerson}({2008})}]{Kamenev2008}
\bibinfo{author}{\bibfnamefont{A.}~\bibnamefont{Kamenev}} \bibnamefont{and}
  \bibinfo{author}{\bibfnamefont{B.}~\bibnamefont{Meerson}},
  \bibinfo{journal}{{Phys. Rev. E}} \textbf{\bibinfo{volume}{{77}}},
  \bibinfo{pages}{{061107}} (\bibinfo{year}{{2008}}).

\bibitem[{\citenamefont{Anderson and May}(1991)}]{AMbook}
\bibinfo{author}{\bibfnamefont{R.~M.} \bibnamefont{Anderson}} \bibnamefont{and}
  \bibinfo{author}{\bibfnamefont{R.~M.} \bibnamefont{May}},
  \emph{\bibinfo{title}{Infectious Diseases of Humans: Dynamics and Control}}
  (\bibinfo{publisher}{Oxford Science Publications}, \bibinfo{year}{1991}).

\bibitem[{\citenamefont{d'Onofrio}(2002)}]{d'Onofrio2002}
\bibinfo{author}{\bibfnamefont{A.}~\bibnamefont{d'Onofrio}},
  \bibinfo{journal}{Math. Biosci.} \textbf{\bibinfo{volume}{179}},
  \bibinfo{pages}{57} (\bibinfo{year}{2002}).

\bibitem[{\citenamefont{Gao et~al.}(2007)\citenamefont{Gao, Chen, and
  Teng}}]{Gao2007}
\bibinfo{author}{\bibfnamefont{S.}~\bibnamefont{Gao}},
  \bibinfo{author}{\bibfnamefont{L.}~\bibnamefont{Chen}}, \bibnamefont{and}
  \bibinfo{author}{\bibfnamefont{Z.}~\bibnamefont{Teng}},
  \bibinfo{journal}{Bull. Math. Biol.} \textbf{\bibinfo{volume}{69}},
  \bibinfo{pages}{731} (\bibinfo{year}{2007}).

\bibitem[{\citenamefont{Shulgin et~al.}(1998)\citenamefont{Shulgin, Stone, and
  Agur}}]{Shulgin1998}
\bibinfo{author}{\bibfnamefont{B.}~\bibnamefont{Shulgin}},
  \bibinfo{author}{\bibfnamefont{L.}~\bibnamefont{Stone}}, \bibnamefont{and}
  \bibinfo{author}{\bibfnamefont{Z.}~\bibnamefont{Agur}},
  \bibinfo{journal}{Bull. Math. Biol.} \textbf{\bibinfo{volume}{60}},
  \bibinfo{pages}{1123} (\bibinfo{year}{1998}).

\bibitem[{\citenamefont{Stone et~al.}(2000)\citenamefont{Stone, Shulgin, and
  Agur}}]{Stone2000}
\bibinfo{author}{\bibfnamefont{L.}~\bibnamefont{Stone}},
  \bibinfo{author}{\bibfnamefont{B.}~\bibnamefont{Shulgin}}, \bibnamefont{and}
  \bibinfo{author}{\bibfnamefont{Z.}~\bibnamefont{Agur}},
  \bibinfo{journal}{Math. and Computer Modelling}
  \textbf{\bibinfo{volume}{31}}, \bibinfo{pages}{207} (\bibinfo{year}{2000}).

\bibitem[{\citenamefont{Wang et~al.}(2009)\citenamefont{Wang, Tao, and
  Song}}]{wang2009}
\bibinfo{author}{\bibfnamefont{X.}~\bibnamefont{Wang}},
  \bibinfo{author}{\bibfnamefont{Y.}~\bibnamefont{Tao}}, \bibnamefont{and}
  \bibinfo{author}{\bibfnamefont{X.}~\bibnamefont{Song}},
  \bibinfo{journal}{Appl. Math. and Computation}
  \textbf{\bibinfo{volume}{210}}, \bibinfo{pages}{398} (\bibinfo{year}{2009}).

\bibitem[{\citenamefont{Dykman et~al.}({2008})\citenamefont{Dykman, Schwartz,
  and Landsman}}]{dykmanschwartz08}
\bibinfo{author}{\bibfnamefont{M.~I.} \bibnamefont{Dykman}},
  \bibinfo{author}{\bibfnamefont{I.~B.} \bibnamefont{Schwartz}},
  \bibnamefont{and} \bibinfo{author}{\bibfnamefont{A.~S.}
  \bibnamefont{Landsman}}, \bibinfo{journal}{{Phys. Rev. Lett.}}
  \textbf{\bibinfo{volume}{{101}}}, \bibinfo{pages}{078101}
  (\bibinfo{year}{{2008}}).

\bibitem[{\citenamefont{Pastor-Satorras and
  Vespignani}(2002)}]{Pastor-Satorras2002}
\bibinfo{author}{\bibfnamefont{R.}~\bibnamefont{Pastor-Satorras}}
  \bibnamefont{and}
  \bibinfo{author}{\bibfnamefont{A.}~\bibnamefont{Vespignani}},
  \bibinfo{journal}{Phys. Rev. E} \textbf{\bibinfo{volume}{65}},
  \bibinfo{pages}{036104} (\bibinfo{year}{2002}).

\bibitem[{\citenamefont{Dezs\H{o} and Barab\'{a}si}(2002)}]{Dezso2002}
\bibinfo{author}{\bibfnamefont{Z.}~\bibnamefont{Dezs\H{o}}} \bibnamefont{and}
  \bibinfo{author}{\bibfnamefont{A.-L.} \bibnamefont{Barab\'{a}si}},
  \bibinfo{journal}{Phys. Rev. E} \textbf{\bibinfo{volume}{65}},
  \bibinfo{pages}{055103(R)} (\bibinfo{year}{2002}).

\bibitem[{\citenamefont{Zanette and Kuperman}(2002)}]{Zanette2002}
\bibinfo{author}{\bibfnamefont{D.~H.} \bibnamefont{Zanette}} \bibnamefont{and}
  \bibinfo{author}{\bibfnamefont{M.}~\bibnamefont{Kuperman}},
  \bibinfo{journal}{Physica A} \textbf{\bibinfo{volume}{309}},
  \bibinfo{pages}{445 } (\bibinfo{year}{2002}).

\bibitem[{\citenamefont{Miller and Hyman}(2007)}]{Miller2007}
\bibinfo{author}{\bibfnamefont{J.}~\bibnamefont{Miller}} \bibnamefont{and}
  \bibinfo{author}{\bibfnamefont{J.~M.} \bibnamefont{Hyman}},
  \bibinfo{journal}{Physica A} \textbf{\bibinfo{volume}{386}},
  \bibinfo{pages}{780} (\bibinfo{year}{2007}).

\bibitem[{\citenamefont{Cohen et~al.}(2003)\citenamefont{Cohen, Havlin, and
  ben-Avraham}}]{Cohen2003}
\bibinfo{author}{\bibfnamefont{R.}~\bibnamefont{Cohen}},
  \bibinfo{author}{\bibfnamefont{S.}~\bibnamefont{Havlin}}, \bibnamefont{and}
  \bibinfo{author}{\bibfnamefont{D.}~\bibnamefont{Ben-Avraham}},
  \bibinfo{journal}{Phys. Rev. Lett.} \textbf{\bibinfo{volume}{91}},
  \bibinfo{pages}{247901} (\bibinfo{year}{2003}).

\bibitem[{\citenamefont{Chen et~al.}(2008)\citenamefont{Chen, Paul, Havlin,
  Liljeros, and Stanley}}]{Chen2008}
\bibinfo{author}{\bibfnamefont{Y.}~\bibnamefont{Chen}},
  \bibinfo{author}{\bibfnamefont{G.}~\bibnamefont{Paul}},
  \bibinfo{author}{\bibfnamefont{S.}~\bibnamefont{Havlin}},
  \bibinfo{author}{\bibfnamefont{F.}~\bibnamefont{Liljeros}}, \bibnamefont{and}
  \bibinfo{author}{\bibfnamefont{H.~E.} \bibnamefont{Stanley}},
  \bibinfo{journal}{Phys. Rev. Lett.} \textbf{\bibinfo{volume}{101}},
  \bibinfo{pages}{058701} (\bibinfo{year}{2008}).

\bibitem[{\citenamefont{Santos et~al.}(2008)\citenamefont{Santos, Paes-Sousa,
  da~Silva~Junior, and Victorad}}]{NationalImmDays}
\bibinfo{author}{\bibfnamefont{L.}~\bibnamefont{Santos}},
  \bibinfo{author}{\bibfnamefont{R.}~\bibnamefont{Paes-Sousa}},
  \bibinfo{author}{\bibfnamefont{J.~B.} \bibnamefont{da~Silva~Junior}},
  \bibnamefont{and} \bibinfo{author}{\bibfnamefont{C.~G.}
  \bibnamefont{Victorad}}, \bibinfo{journal}{Bull World Health Organ.}
  \textbf{\bibinfo{volume}{86}}, \bibinfo{pages}{474} (\bibinfo{year}{2008}).

\bibitem[{not({\natexlab{a}})}]{note1}
\bibinfo{note}{Steady state behavior occurs for other $q$ values, but we focus
  here on the oscillatory regions because the dynamics is more interesting. The
  qualitative advantage to including both vaccination and rewiring is the same
  whether or not oscillations occur.}

\bibitem[{not({\natexlab{b}})}]{note2}
\bibinfo{note}{The bifurcation structure and infective levels depend
  primarily on the average vaccination rate $\nu A$ rather than on the
  frequency or amplitude individually.}

\end{thebibliography}
\end{document}